\documentclass[aps,prl,floatfix,showpacs,twocolumn]{revtex4}
\usepackage{amssymb,amsmath}
\usepackage{stmaryrd}
\usepackage{graphicx}
\usepackage{epsfig}
\usepackage{psfrag}

\begin{document}

\hsize\textwidth\columnwidth\hsize\csname@twocolumnfalse\endcsname
\title{Quantized Casimir Force} 

\author{Wang-Kong Tse}
\author{A.~H. MacDonald}
\affiliation{Department of Physics, University of Texas, Austin, Texas 78712, USA}

\begin{abstract}
We 
investigate the Casimir effect between two-dimensional electron 
systems driven to the quantum Hall regime by a strong perpendicular magnetic 
field. In the large separation ($d$) limit where retardation effects are essential
we find i) that the Casimir force is quantized in units of 
${3\hbar c}\alpha^2/{8\pi^2}{d^4}$, and ii) that the 
force is repulsive for mirrors with same type of carrier, 
and attractive for mirrors with opposite types of carrier. 
The sign of the Casimir force is therefore electrically tunable in ambipolar 
materials like graphene.  The Casimir force is 
suppressed when one mirror is a charge-neutral graphene system 
in a filling factor $\nu=0$ quantum Hall state. 
\end{abstract}
\maketitle

\newpage




\textit{Introduction ---} 
The Casimir effect is an intriguing quantum electrodynamic phenomenon
in which the vacuum energy of the electromagnetic field 
is altered by the presence of two closely-spaced mirrors coupled to
the field,
resulting in a measurable force between 
them. The force between two parallel metal plates, for 
example, is attractive because the plates 
restrict the number of vacuum electromagnetic modes present in the 
space between them. 
Growing interest in this facscinating topic has been driven by improving
Casimir force measurement capabilities
at small separations from $1\,\mu$m down to about $10$ nm
\cite{
Mohi_exp1,Eder_exp1,Capasso_exp2,Decca_exp1} and by new materials.
When semiconductors like Si are used as mirrors 
instead of conventional dielectric materials and metals, for example,  
it is possible to control the Casimir force by 
optical or electrical carrier density modulation \cite{Casimir_RMP}.
This property, combined with recent advances in micro-electromechanical systems (MEMS) and
nano-electromechanical systems (NEMS), could enable new
electromechanical applications. 

At submicron distances, the Casimir
force between materials usually dominates the gravitational force. 
Measurement of the Casimir force
has therefore played an important role in the search for the  
non-Newtonian gravitational forces suggested by a number of
unification theories \cite{unify}.
Because the Newtonian gravitational law has been poorly tested below $10\,\mu\mathrm{m}$, 
an accurate quantitative understanding of the Casimir effect is key to
the identification of any new gravitational force law which might  
prevail at small length scales.
Theories of the Casimir effect have in fact  
been employed \cite{gravityprop,gravityexp}
in establishing constraints on the magnitude of non-Newtonian 
gravitational effects.  In these studies measured force values are
compared with known contributions from Newtonian gravity, 
and Casimir forces computed using a separate set of
measurements of the detailed dielectric properties of 
the test materials \cite{Advances}. The Casimir force can be sensitive to 
unintended variations in impurity density, degree of surface roughness, and
sample thickness.
The errors associated with Casimir force estimation are therefore difficult to control 
and this in turn crucially 
limits the precision of bounds placed on hypothetical 
gravitational forces. 
A way to suppress or even neutralize 
the Casimir effect at small length 
scales is therefore desirable, enabling a more sensitive direct measurement of gravity. 

In this Letter, we address one possibility for suppressing 
the Casimir effect 
by developing
a theory of the Casimir force between two-dimensional (2D) electron
systems in the presence of strong perpendicular
magnetic fields. The development of quantized Landau levels and the
associated quantum Hall effect opens up a new regime for investigations of the 
Casimir effect. We show that in ambipolar materials 
like graphene the Casimir force
is electrically tunable between attractive and repulsive values, and that 
in the large-separation relativistic
regime, the Casimir force is quantized. 
Importantly, a 
strongly suppressed Casimir force can be achieved under a high magnetic field by using charge-neutral
graphene sheets as mirrors. 

\textit{Theory ---} When spatial dispersion in the mirrors is negligible,
the Casimir effect is determined by their local 
(\textit{i.e.}, 
$q = 0$) 
charge and current response
functions. 
In this limit, the Casimir energy (per unit area) can be elegantly expressed
in terms of the reflection coefficients of the mirrors \cite{Reynaud} 
%
\begin{eqnarray}
E &=&
\frac{\hbar}{4\pi^2}\int_0^{\infty}\mathrm{d}q_{\perp}q_{\perp}\int_0^{q_{\perp}c}\mathrm{d}\omega
\nonumber \\
&&\mathrm{tr}\,\mathrm{ln}
\left[\mathbb{I}-r'^{\mathrm{L}}(iq_{\perp},i\omega)r^{\mathrm{R}}(iq_{\perp},i\omega)\right], \label{Lif}
\end{eqnarray}
where $r'^{\mathrm{L}}, r^{\mathrm{R}}$ are the reflection 
coefficients \textit{into} the cavity of the left (L) and right
(R) mirrors, $q_{\perp} = \sqrt{(\omega/c)^2-q^2}$ is the component
of photon momentum perpendicular to the mirror planes, and 
$q$ the in-plane momentum component.  

We apply Eq.~(\ref{Lif}) to mirrors made of ultrathin films 
that can be adequately modeled as a quasi-2D layer; this 
class of systems includes atomically thin materials like 
graphene and bilayer graphene \cite{GraphRev}. Several groups have
studied the Casimir force between graphene sheets in the 
absence of an external
magnetic field \cite{graphene_Casimir}.
%
When a field is present the optical characteristics of a 2D mirror depend on its
longitudinal $\sigma_{xx}$ and Hall $\sigma_{xy}$
conductivities.  For general 
angle of incidence $\theta$, straightforward calculations yield \cite{TI_Mag} the 
following expressions for the tensor components of the reflection coefficients of a 2D 
mirror in a plane with fixed $\hat{z}$ direction coordinate $z$:
\begin{eqnarray}
r_{xx} &=& -2\pi e^{i2q_{\perp}
z}\left\{\bar{\sigma}_{xx}(\omega)/\lambda+2\pi\left[\bar{\sigma}_{xx}^2 (\omega)+\bar{\sigma}_{xy}^2 (\omega)\right]\right\}/{R},\nonumber \\
r_{xy} &=& -r_{yx} = -2\pi e^{i2q_{\perp} z}\bar{\sigma}_{xy}(\omega)/R, \nonumber \\
r_{yy} &=& -2\pi e^{i2q_{\perp}
z}\left\{\bar{\sigma}_{xx}(\omega)\lambda+2\pi\left[\bar{\sigma}_{xx}^2(\omega)+\bar{\sigma}_{xy}^2(\omega)\right]\right\}/{R},
\nonumber \\
\label{SLRefl}
\end{eqnarray}
%
where $\lambda = \mathrm{cos}\theta = q_{\perp}/(\omega/c)$, ${R} =
1+2\pi\bar{\sigma}_{xx}(\lambda+1/\lambda)+4\pi^2[\bar{\sigma}_{xx}^2+\bar{\sigma}_{xy}^2]$, 
and
$\bar{\sigma}_{xx,xy} = \sigma_{xx,xy}/c$ are \textit{dimensionless}
optical conductivities normalized by $c$. The expression for $r'$ differs 
by the replacement 
$\mathrm{exp}(i2q_{\perp} z) \to \mathrm{exp}(-i2q_{\perp} z)$. 

We now apply a strong magnetic field normal to two parallel 2D mirrors 
that are separated by a distance $d$. 
When $d/c$ is larger than  
characteristic electronic time scales the  
retardation effects captured by Eq.~\ref{Lif} are essential.  
In the quantum Hall regime, the large $d$ limit holds for $d/c$ much larger than 
$\mathrm{max}(\hbar/\Delta,\hbar/\Gamma)$ where $\Delta$ describes 
an inter-Landau level transition energy and $\Gamma$ a disorder broadening energy.  
%
Because $r'^{\mathrm{L}}r^{\mathrm{R}} \propto \mathrm{exp}(-2q_{\perp}d)$ in 
Eq.~(\ref{Lif}), electromagnetic correlations between mirrors
separated by large distances are carried by long-wavelength virtual
photons; the integral in Eq.~(\ref{Lif}) is therefore  
dominated by low-frequency contributions. The value of 
the Casimir energy is thus determined at large separations by the static longitudinal
and Hall conductivities
$\sigma_{xx,xy}(\omega = 0)$.
If the applied magnetic field is strong enough that the 
mirrors sit 
on well-formed quantum Hall plateaus,
then the low-frequency longitudinal conductivity vanishes 
$\bar{\sigma}_{xx}^{\mathrm{L,R}} \simeq 0$ and the 
Hall conductivity $\bar{\sigma}_{xy}^{\mathrm{L,R}} \simeq
\nu_{\mathrm{L,R}}\alpha/2\pi$, where $\alpha = 1/137$ is the fine
structure constant and $\nu_{\mathrm{L,R}} = 2\pi \hbar
n_{\mathrm{L,R}}/(eB)$ are the Landau level filling factors of the left and 
right mirrors with carrier density $n_{\mathrm{L,R}}$. 
Expanding the reflection coefficients Eq.~(\ref{SLRefl}) up to leading order in 
$\alpha$ yields $\mathrm{tr}\,\mathrm{ln}[\mathbb{I}-r'^{\mathrm{L}}r^{\mathrm{R}}]
\simeq
\mathrm{ln}(1+2 e^{-2q_{\perp}d}\nu_{\mathrm{L}}\nu_{\mathrm{R}}\alpha^2)
\simeq 2 e^{-2q_{\perp}d} \nu_{\mathrm{L}}\nu_{\mathrm{R}}\alpha^2$.
It then follows from Eq.~(\ref{Lif}) that the 
Casimir force per unit area $F = -\partial E/\partial d$ 
between two quantum Hall mirrors is 
\begin{equation}
F =
\frac{3\hbar c}{8\pi^2}\alpha^2\frac{\nu_{\mathrm{L}}\nu_{\mathrm{R}}}{d^4}. \label{Main}
\end{equation}
Eq.~(\ref{Main}) is 
a central result of this Letter.  Because the quantum Hall effect occurs 
only for discrete integer and small-denominator rational values of $\nu$,
we conclude that the large-separation asymptotic 
Casimir force is quantized in units of ${3\hbar
  c}\alpha^2/{8\pi^2}{d^4}$. Comparing with the well-known result
between perfect metals $F_0 = -\hbar c\pi^2/240 d^4$, 
one notices that
the 
Casimir force is suppressed by a factor $\propto \alpha^2$.

%
\begin{figure}
  \includegraphics[width=8.5cm,angle=0]{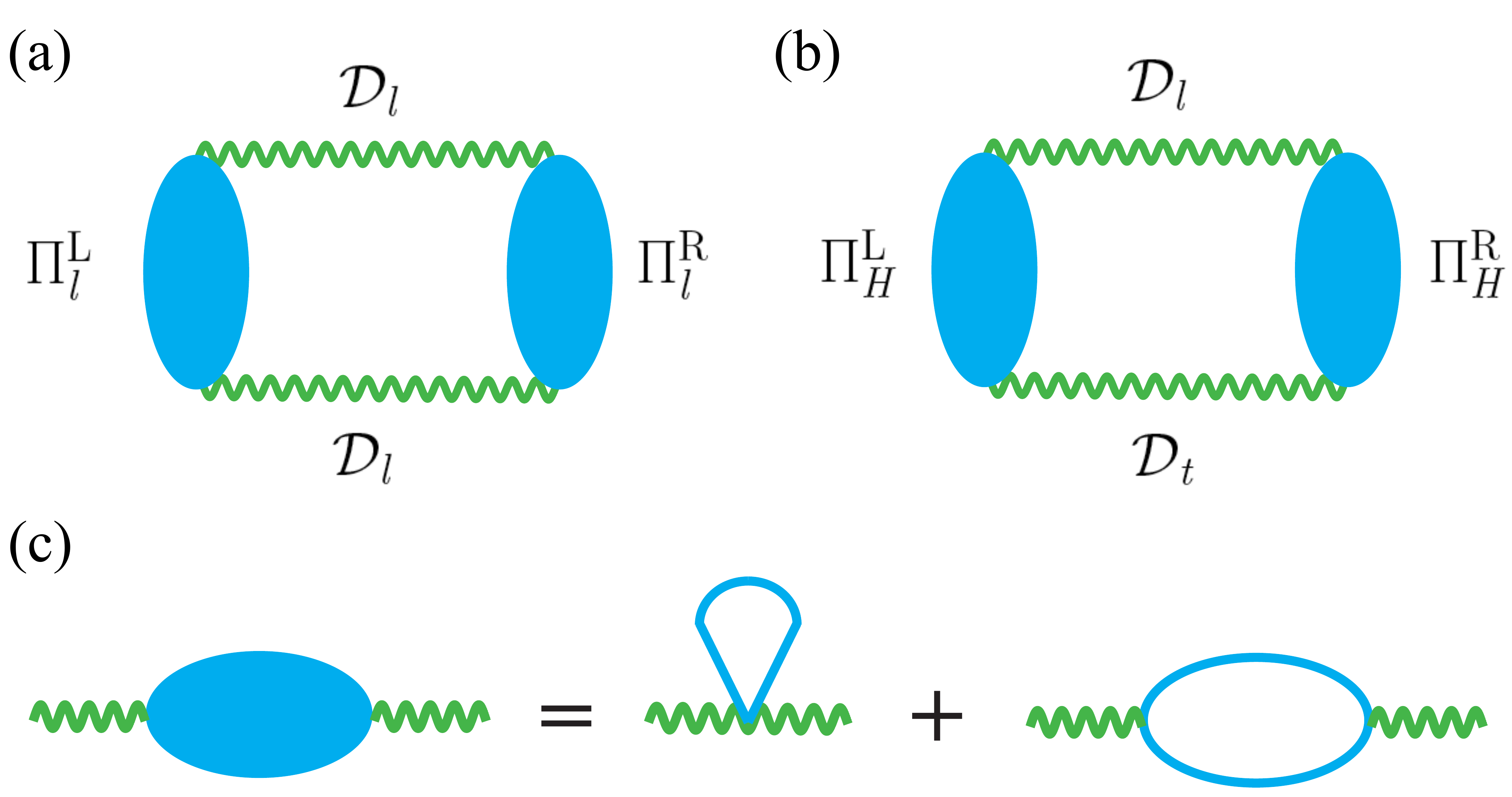}
\caption{(Color online). Feynman diagrams for the ground state
  correlation energy between mirrors, where green wavy lines represent the photon Green
  function $\mathcal{D}$ and filled blue/dark bubbles represent the current-current
  correlation function $\Pi$. (a). Leading-order diagram for the
  longitudinal contribution $E_{{l}}$ to the ground state correlation
  energy. The transverse contribution $E_{{t}}$ is given by  
  replacing $\mathcal{D}_{{l}},\Pi_{l}\,\to\,\mathcal{D}_{{t}},\Pi_{t}$. (b) Leading-order diagram for the 
  Hall contribution $E_{\mathit{H}}$, this diagram is to be counted
  twice because an equivalent diagram results from interchanging
  $\mathcal{D}_{{l}}$ and $\mathcal{D}_{{t}}$ and 
  using $\Pi_{\mathit{H}} \equiv \Pi_{xy} = -\Pi_{yx}$. (c) Diagram for the
  current-current correlation function. The first diagram on the right 
represents the diamagnetic contribution whereas the second represents
the paramagnetic contribution.} 
\label{Fey1}
\end{figure}

It is informative to also understand Eq.~(\ref{Main}) from the perspective of
quantum electrodynamic perturbation theory in which the  
Casimir effect arises from mutual electromagnetic correlations 
between two separated systems and therefore contributes to the 
ground-state \cite{APReview,Pit,Barash} energy of the coupled system. 
In the following, we adopt the axial gauge with a zero 
electromagnetic scalar potential $\phi = 0$.  The longitudinal (photon
momentum parallel to vector potential) $\mathcal{D}_{l}$ and transverse
(photon momentum perpendicular to vector potential) $\mathcal{D}_{t}$ 
interlayer photon propagators \cite{Marusic_Propag} are given as a function of mirror separation by 
$\mathcal{D}_{l}(q_{\perp},\omega) = -(2\pi i c 
q_{\perp}/\omega^2)\mathrm{exp}(iq_{\perp}d)$ and $\mathcal{D}_{t}(q_{\perp},\omega) = -[2\pi i
/(q_{\perp}c)]\mathrm{exp}(iq_{\perp}d)$. In the absence of a magnetic
field, the longitudinal contribution to the interaction energy is given to leading order in $\alpha$ by 
the diagram in Fig.~\ref{Fey1}a:
%
%
\begin{equation}
E_{{l}} =
-\hbar\int\frac{\mathrm{d}\omega}{2\pi}\sum_q\Pi_{l}^{\mathrm{L}}\mathcal{D}_{l}\Pi_{l}^{\mathrm{R}}\mathcal{D}_{l}, \label{El} 
\end{equation}
where $\Pi_{l}^{\mathrm{L,R}}$ is the longitudinal component of the
current-current correlation tensor of the mirrors. 
By virtue of the continuity equation for charge density,
the longitudinal current-current correlation function is related to the 
density-density correlation function $\chi$ by $\Pi_{l}(q,\omega) = 
(e\omega/q)^2\chi(q,\omega)$.
Eq.~(\ref{El}) therefore describes the
leading-order contribution to the Casimir effect due to fluctuations
of charge density. 
Because $\Pi_{l} \propto \sigma_{xx}$ is independent of the sign of 
charge carriers, the Casimir force between two
parallel-plate mirrors placed in vacuum is attractive 
for like or unlike carrier densites. The transverse interaction energy
$E_{t}$ gives a comparable contribution to Eq.~(\ref{El}) with integrand
$\Pi_{t}^{\mathrm{L}}\mathcal{D}_{t}\Pi_{t}^{\mathrm{R}}\mathcal{D}_{t}$
where  $\Pi_{t}^{\mathrm{L,R}}$ is the transverse component of the
current-current correlation tensor.


When a magnetic field is present, the Casimir 
energy acquires a contribution that depends on the Hall current-current correlation function 
$\Pi_{\mathit{H}}(q,\omega)$, for which the leading-order diagram (Fig.~\ref{Fey1}b) yields
\begin{equation}
E_{\mathit{H}} = 
 2\hbar\int\frac{\mathrm{d}\omega}{2\pi}\sum_q\Pi_{\mathit{H}}^{\mathrm{L}}\mathcal{D}_{t}\Pi_{\mathit{H}}^{\mathrm{R}}\mathcal{D}_{l}. \label{Et}
\end{equation} 
Expressed in terms of the Hall conductivity, $\Pi_{\mathit{H}} = -i\omega\sigma_{xy}$ in
Eq.~(\ref{Et}).  Performing the complex plane rotations  \cite{Reynaud} $q_{\perp}
\to iq_{\perp}$ and $\omega \to i\omega$, we therefore obtain  
%
\begin{equation}
E_{\mathit{H}} =
\frac{2\hbar}{c^2}\int_0^{\infty}\mathrm{d}q_{\perp}q_{\perp}\int_0^{q_{\perp}c}{\mathrm{d}\omega}\sigma_{xy}^{\mathrm{L}}(iq_{\perp},i\omega)\sigma_{xy}^{\mathrm{R}}(iq_{\perp},i\omega)^{-2q_{\perp}d}, \label{Etcond}
\end{equation}
showing that the Hall contribution $E_{\mathit{H}}$ depends on the product of the Hall conductivities of 
the two mirrors, and is therefore sensitive to the signs of their
charge carriers. To leading order in $\alpha$,
the Casimir energy in a magnetic field is given by the sum of these
three contributions $E = E_{{l}}+E_{{t}}+E_{\mathit{H}}$.  

When the magnetic field is strong, Landau quantization of the electronic energy
spectrum of the mirror leads to energy gaps $\Delta$. 
We confine our discussion to
low temperatures $k_{\mathrm{B}}T \ll \hbar c/d \ll \Delta$ at which the quantum
Hall effect is established and thermal contributions to the Casimir
force can be ignored. 
In this limit density fluctuations and hence the longitudinal and 
transverse contributions to the Casimir energy
are suppressed because the 2D quantum Hall
systems are incompressible \cite{Note1}.
Hence, Hall-like quantum fluctuations of the current dominates. 
Evaluating the Hall energy in Eq.~(\ref{Etcond}), 
we can rederive 
the Casimir force result Eq.~(\ref{Main}) 
obtained earlier from macroscopic electrodynamic considerations. 
\begin{figure}
  \includegraphics[width=8.7cm,angle=0]{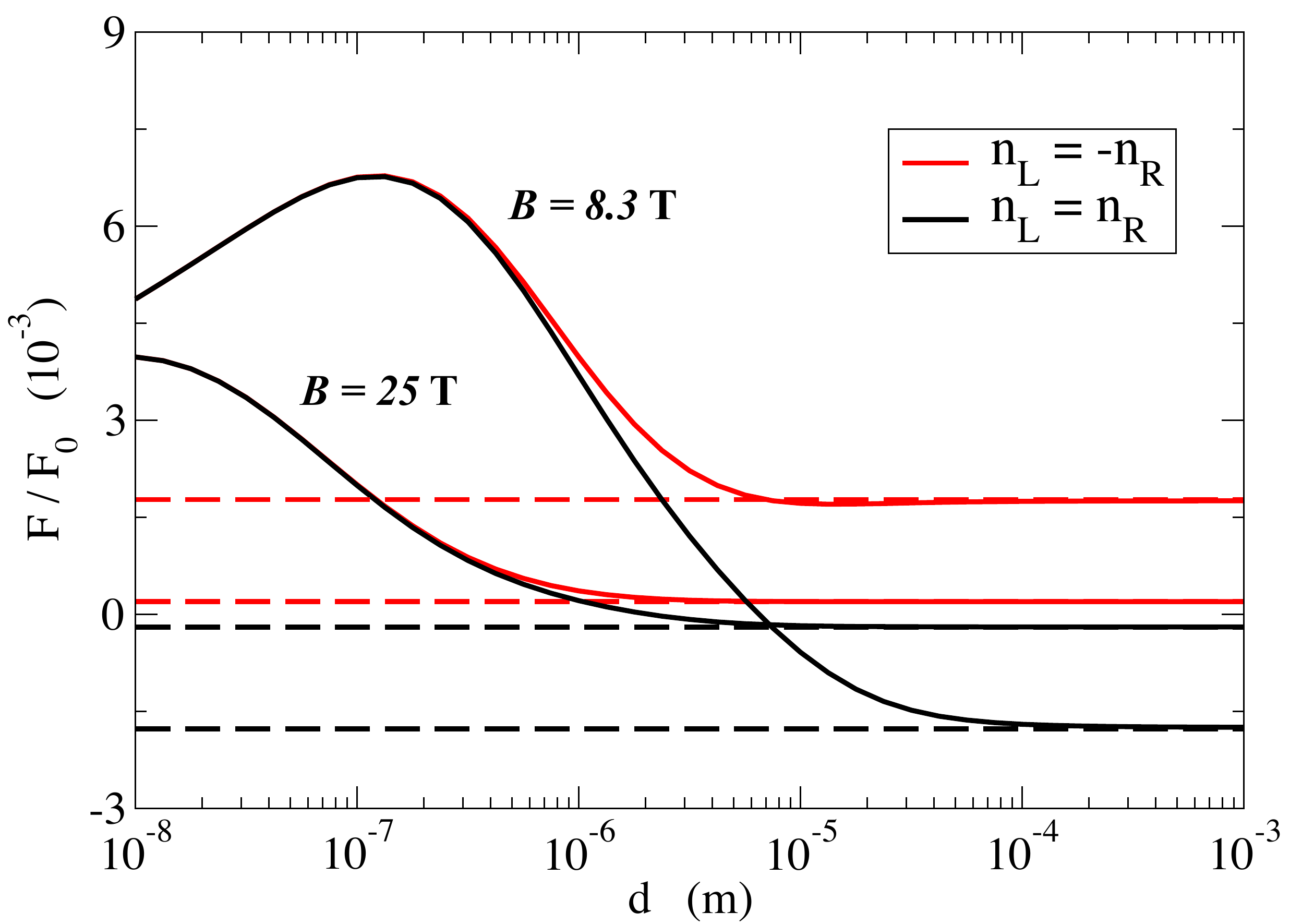}
\caption{(Color online). Casimir force $F$ normalized by the perfect 
  metal value $F_0 = -\hbar c \pi^2/240 d^4$ as a function of
  separation $d$ in a strong magnetic field. Solid lines correspond to numerical results and
  dashed lines to the analytic large $d$ limits in Eq.~(\ref{Main}). The case with
  opposite-sign carrier densities on the two graphene sheets $n_{\mathrm{L}} =
  3\times10^{11}\,\mathrm{cm}^{-2} = -n_{\mathrm{R}}$ is depicted by
  the red (grey) curve whereas the case with the same-sign carrier density 
  $n_{\mathrm{L}}=n_{\mathrm{R}}=3\times10^{11}\,\mathrm{cm}^{-2}$ is depicted by the black (dark)
  curve.  At these carrier densities, the magnetic fields
  $B = 25, 8.3\,\mathrm{T}$ correspond to filling factors $\vert \nu\vert = 2, 6$.
  These calculations are valid for $d \gg l_{B}$ where 
$l_{B} = 257/\sqrt{B}\,{\AA}$ 
is the magnetic length
  and weak disorder.} 
\label{Plot1}
\end{figure}

\textit{Repulsive Casimir Effect ---} Repulsive Casimir forces are unusual and have been 
realized for the first time in recent experiments 
involving test bodies immersed in a liquid
medium \cite{Capasso_exp1}. 
An important implication of Eq.~(\ref{Main}) is that repulsive
Casimir forces can be easily obtained in ambipolar 2D quantum 
Hall systems by gating them to like-sign 
carrier 
densities.  
We demonstrate this repulsive Casimir effect by considering the 
case of two 
graphene sheets, numerically evaluating the Casimir
force $F = -\partial 
E/\partial d$ 
from the Casimir energy $E$ given by Eqs.~(\ref{Lif})-(\ref{SLRefl}).
We obtain the graphene sheet's optical conductivity tensor 
from the Kubo formula \cite{Suppl}. 
\begin{figure}
  \includegraphics[width=8.7cm,angle=0]{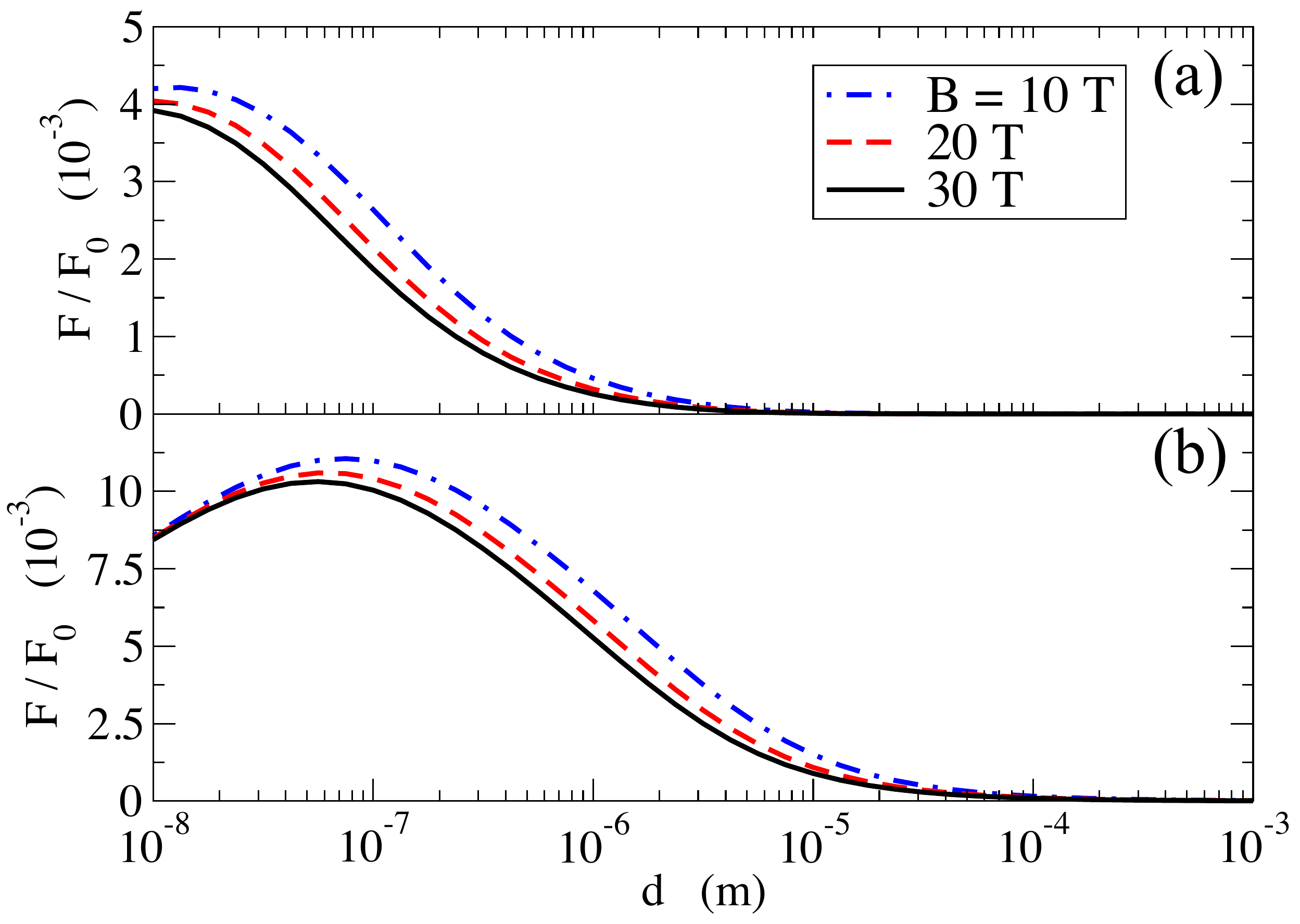}
\caption{(Color online). Casimir force versus separation for magnetic field strengths $B = 10,
  20, 30\,\mathrm{T}$ between (a) two charge-neutral graphene sheets;
  (b) a charge-neutral graphene sheet and a half-space composed of an
  insulator with dielectric constant $\epsilon$. In the large $d$ regime where the 
  suppression effect becomes prevalent, the insulator is characterized
  by its low-frequency dielectric behavior, and we take $\epsilon \simeq 4$ corresponding to
  SiO$_2$ or hexagonal BN.
} 
\label{nu0}
\end{figure}
Fig.~\ref{Plot1} shows the calculated Casimir force 
versus $d$ between (1) an electron-doped graphene sheet and a hole-doped graphene sheet 
(grey/red curve), and (2) between two electron-doped graphene sheets
(dark/black curve). The plot is normalized by the Casimir force
between two perfect metals $F_0 = -\hbar c \pi^2/240 d^3$. 
For small separations the Casimir force is attractive 
for both cases.
For separations large enough that retardation effects are important 
however, contributions from charge density fluctuations 
are exponentially suppressed, Hall
current fluctuations gain dominance, and repulsive forces can appear.  
Our numerical results for both cases settle 
onto the values predicted by the analytic result Eq.~(\ref{Main}) beyond a
threshold separation, $d_{\mathrm{QH}}$.
For $d > d_{\mathrm{QH}}$ we find Casimir attraction for 
opposite-sign carrier densities and repulsion for like-sign
carrier densities. 
The crossover length $d_{\mathrm{QH}}$ can be estimated 
by expanding Eq.~(\ref{Lif}) to one higher-order term in 
$1/d$ and $\alpha$ beyond the limit given by Eq.~(\ref{Main}).
We find \cite{Suppl} that $d_{\mathrm{QH}} \gtrsim 
l_B(c/v)f(\nu_{\mathrm{L}},\nu_{\mathrm{R}})$, where $l_B =
\sqrt{\hbar/eB}$ is the magnetic length, $v = 10^6\,\mathrm{ms}^{-1}$ is the band velocity in graphene, 
and $f(\nu_{\mathrm{L}},\nu_{\mathrm{R}})$ is a monotonic increasing
function of $\nu_{\mathrm{L,R}} \propto 1/B$ of order $\mathcal{O}(1)$. 
In agreement with numerical results in Fig.~\ref{Plot1}, $d_{\mathrm{QH}}$
decreases with magnetic field. 

\textit{Casimir Force Quenching ---} 
We have shown that the asymptotic Casimir force between 2D quantum Hall insulators follows the same power 
law as the Casimir force for ideal thick metals but is weaker by a 
factor  of $\pi^4 \nu_{L} \nu_{R}  \alpha^2/90$ 
where $\alpha$ is the fine structure constant.   At the same time it is qualitatively stronger than 
the Casimir force between ordinary 2D insulators which falls off especially
rapidly (like $d^{-6}$) at large distances. 
For the special case of charge-neutral graphene in a $\nu=0$ quantum
Hall state \cite{nuzeroQH}, the leading Hall contribution Eq.~(\ref{Main}) to the Casimir force 
vanishes, 
and the system behaves like an ordinary 2D insulator, with the Casimir force
given asymptotically at large $d$ by $F = -9\hbar c\,{\sigma_{xx}^{\mathrm{L}}}'(0)
{\sigma_{xx}^{\mathrm{R}}}'(0)/2d^6$ where $'$ denotes a derivative with
 respect to frequency \cite{Suppl}. 
Fig.~\ref{nu0}(a) shows the 
calculated Casimir forces 
between two parallel charge-neutral graphene sheets as a function of magnetic field strength. 
The force value 
is substantially suppressed 
for 
large 
separations $d \gtrsim 10\,\mu\mathrm{m}$ at a typical 
field strength of $10\,\mathrm{T}$, indicating that the
relativistic-regime Casimir force is quenched by strong magnetic
fields. 
In this regime, we find \cite{Suppl} that 
\begin{equation}
F = -\frac{9\hbar
  c^3}{4\pi^2d^6}\alpha^2\left(\frac{l_B}{v}\right)^2\left[\frac{1}{2}+h(1)\right]^2, \label{F01}
\end{equation}
where $h(n) = \sum_{m = n}^{\infty}{1}/{\left(\sqrt{m}+\sqrt{m+1}\right)^3}$ and $h(1)
\approx 0.25$. For $d \simeq 10-100\,\mu\mathrm{m}$, $F$ is
suppressed by $\sim 10^{-8}-10^{-10}$ compared to the ideal metal value $F_0$.

Because of experimental difficulties  
in maintaining precise parallelism between two plates, it is common in
Casimir force measurements to replace one of the plates by a sphere 
to mimic the parallel-plates geometry at the point of closest
separation. It is therefore essential 
to 
ascertain 
whether the quenching effect of the Casimir force 
also survives under this geometry. 
Fig.~\ref{nu0}(b) shows the full numerical results of the
Casimir force between a charge-neutral graphene sheet and a dielectric
half-space mimicking a large insulating sphere with radius $R \gg d$. 
The Casimir force 
again drops rapidly, and does so at a larger value of $d$ due to the
higher opacity of the dielectric region. The
quenching effect of the Casimir force should therefore remain observable in a 
sphere-plate geometry with a charge-neutral graphene sheet in the $\nu = 0$ quantum
Hall state. 

The strong quantum Hall effects of graphene 2D electron systems might be 
attractive for circumstances where a weak Casimir force is desirable,
such as in experimental tests of the short-distance gravitational law. 
The $\nu=0$ state in particular has the exceptionally weak Casimir 
force characteristic of 2D insulators.  The quantum Hall effect can 
also be used to gate the graphene sheet accurately to charge 
neutrality, ensuring that electrostatic forces
are absent. 

In conclusion, we find that the large-separation Casimir force
asymptote is quantized in the quantum Hall regime, and that 
in ambipolar systems it can be tuned electrically between attractive and repulsive 
values.  When one of the mirrors is 
charge neutral, strong suppression of the 
Casimir effect can be 
achieved. 
Our findings 
suggest that the use of two-dimensional electronically gapped
materials offer a new strategy for control of the Casimir effect. 

The authors thank Ricardo Decca for helpful comments. This work was supported by Welch Foundation grant TBF1473, NRI-SWAN, and DOE Division of Materials Sciences and Engineering grant DE-FG03-02ER45958.

\end{document}